\documentclass{IOS-Book-Article}

\usepackage{mathptmx}
\usepackage{graphicx}
\begin{document}
\begin{frontmatter}              

\title{Detection of Noisy and Flickering Pixels from SWIFT BAT Event Data}
\runningtitle{IOS Press Style Sample}

\author[A]{\fnms{Arkadip Basak}%
\thanks{arkadipbasak2@gmail.com}},

\address[A]{Department of Physics, Birla Institute of Technology and Science - Pilani, Hyderabad Campus, Hyderabad - 500078, Telengana, India.}

\begin{abstract}
This document presents novel algorithms for detection of noisy and flickering pixels from Burst Alert telescope event data and subsequent elimination of data from such pixels to create a filtered event file. The approach adopted for this purpose is quite different from the current practises and focuses more on the temporal variation of data in the detector pixels over long intervals of time against the current  algorithms which follow a pixel based approach.
\end{abstract}


\end{frontmatter}

\thispagestyle{empty}
\pagestyle{empty}

\section*{Introduction}
Detection of Gamma Ray Bursts with BAT \footnote[2]{Burst Alert Telescope} comprises two steps, the detection of the onset of a burst by looking for surges in the event rate across the detector plane followed by the formation of an image of the sky using the events detected during the time interval at the beginning of the burst. However, owing to the presence of noisy and flickering pixels coupled with the presence of high background, the surges in the event rates in the detector plane can be overlooked and prevent detection of short Gamma Ray Bursts which don't produce drastic increases in event rates. The event data when stripped of the contributions from such noisy and flickering pixels can facilitate the detection of shorter Gamma Ray Bursts.


\section{Detection of Noisy pixels}

\subsection{Selection of Ideal Time Interval}
In order to facilitate faster computation, for event data files with exposure times greater that \emph{1 Ks}, the Ideal time interval needs to found out. However, for data sets with exposure time less than \emph{1 Ks}, the detection of flickering and noisy pixels can be carried out on the entire data set. An Ideal time interval may be defined as a time interval during which events have been reported on maximum number of pixels rather than being saturated over a particular cluster of pixels within the detector.

\subsubsection{Algorithm for obtaining Ideal time interval from a given event file.}

\begin{enumerate} \item Segregate the given data set into time intervals of $1Ks$. Once done, we will have $N$ data sets each comprising of event data for $1Ks$.
\item Define $N$ matrices of size $286 \times 173$ (size of the detector\cite{r1}) each. The values of all the elements of the matrices should be assigned to $1$.
\item Scan each data set and once an event on a particular pixel has been encountered, increase the value of the corresponding matrix element by $1$. This process has to be repeated for $N$ data sets with $N$ available matrices.
\item For each matrix find the product of all the elements. The matrix with the maximum value of the product corresponds to the data set contained by the ideal time interval.
\end{enumerate}

\subsection{Implementation of Noise Removal Algorithm}
BAT images have noise characteristics different from most soft X-Ray images. It is worth noting that BAT images possess noise which follow Gaussian statistics\cite{r1}. Once the Ideal time interval for a particular event file has been identified, data from noisy pixels can be eliminated using the following algorithm.

\subsubsection{Algorithm for detection of data from noisy pixels.}

\begin{enumerate} \item Consider the PHA\cite{r1} data contained within the ideal time interval for the given event file. Find the minimum and the maximum value of the PHA column within the given time interval. Find the difference between the minimum and the maximum values. Let us assume this value to be $N$.
\item Create a 2d array of size $N \times 2$. The first column of the array should contain the PHA values in an ascending order while the second one should contain the no of pixels which contain that PHA value within the given time interval for a particular data set. Plot the first column on the X axis and the second one on the Y axis.  This is similar to creating a histogram of the given data with $N$ being the no of bins.
\item Once plotted, we observe a peak following poisson statistics along with $n$ other peaks following Gaussian statistics. As mentioned earlier noise in BAT follow Gaussian statistics. Thus, if we are able to get rid of these n Gaussian peaks, we can eliminate noise from the given data.
\end{enumerate}

The no of bins can be reduced to values around $N/p$ , $p$ being an integer to increase computation speed. Typically $p$ can take values between 2  and 8, without compromising accuracy.

\subsubsection{Algorithm for eliminating data from noisy pixels.}

\begin{enumerate} \item Select a range within the given set of PHA values which contains the n Gaussian peaks. Typical values of range may be from about $1\% - 40\%$ of the maximum PHA value. It is also to be noted noise is mostly contained in the lower PHA channels. Hence it is not advisable to include very high PHA values in the range as it will lead to an increase in computation time which is undesirable.
\item Scan the array created in the previous section within the given range of PHA values starting from the maximum value of the given range. Typically, the maximum value of the range lies somewhere in the data which follow poisson statistics hence the value in the second column must decrease as we move towards the other end of the range.
\item While scanning if a value is encountered which has more than $x \times f$ pixels associated with it where f is the no of pixels associated with the previous value, then it can be considered as a beginning of a Gaussian peak. Typical values $x$ lies between 1.01 and 1.6 depending upon the no of bins and the distribution of data.
\item The end of the peak occurs when a value less than f has been encountered in the second column of the array. All the values between the beginning and end are flagged as noise and the PHA column in the data set is assigned with a dummy value $D$ for all such values. $D$ shouldn't be any value present in the PHA column of the data set.
\item This process is repeated $n$ times until $n$ Gaussian peaks have been removed.
\end{enumerate}

\begin{figure}[htp]
\centering
\includegraphics[width=11cm]{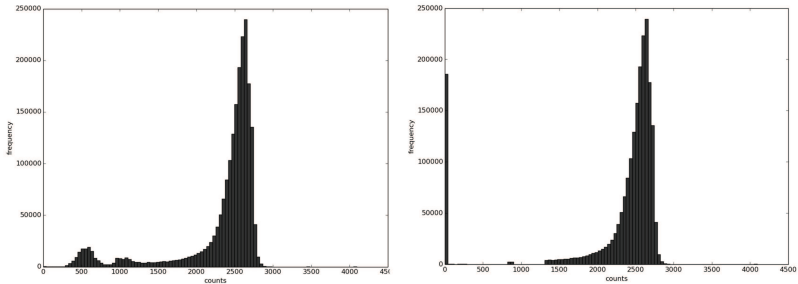}
\caption{Histograms before and after the noise removal algorithm has been applied with $N = 500$ and $D = 0$}
\label{fig:lion1}
\end{figure}

\section{Detection of Flickering Pixels}
Pixels flicker at various timescales which make it difficult to detect \cite{r2}.Thus flickering pixels can be detected at three timescales : fast ($0.1s$), medium ($1s$) and slow ($10s$). At a given timescale, the following algorithm is applied. 
\subsection{Algorithm for detecting flickering pixels}
\begin{enumerate} \item Extract PHA v/s time\footnote[2]{The values in the data set which have been identified as data from noisy pixels need not be extracted as they have already been assigned with a dummy value.} for a single pixel from a particular data set.
\item Bin the data obtained at the selected timescale: : fast ($0.1s$), medium ($1s$) and slow ($10s$). This gives us $N$ independent bins. 
\item The mean ($\mu$) and the standard deviation ($\sigma$) of the bins is found out. 
\item Select a threshold $\mu \pm n\times\sigma$ depending upon the timescale. Typically the value of $n$ can vary between 4 and 10 depending upon the timescale the flickering pixels are to be detected in.
\item Define outlier bins as the ones which contains PHA values outside the given threshold.
\item Sum up the values of the outlier bins. If this sum exceeds 1\% of the sum of data in all the bins, the pixel must be flagged as flickering.
\item Once a pixel is flagged as flickering, assign the contributions from that pixel to the data set with a Dummy value ($D$). This process is repeated for all the pixels.
\end{enumerate}

\section*{Results}
The given algorithm has been applied to a data set titled $sw00030352069bevshpo-uf.evt$\cite{r3} and the results have been plotted below. 

\begin{figure}[htp]
\centering
\includegraphics[width=11cm]{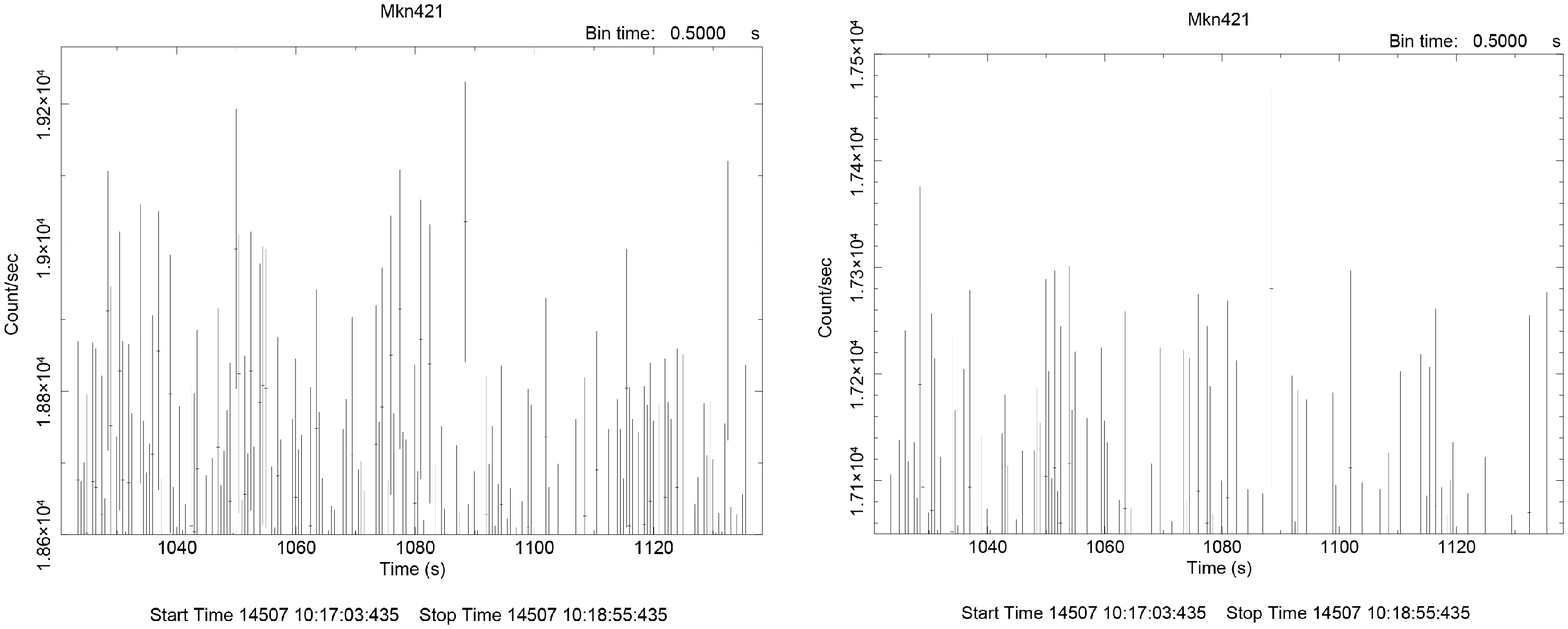}
\caption{Light curves of the data before and after the algorithm has been applied with bin time = 0.5s}
\label{fig:lion}
\end{figure}

On comparing the two plots we observe that there has been a decrease in the global count rate. Also on comparing the peaks present in  both the figures, the peaks at $\approx$ 1017s and $\approx$ 1089s remains while others are no longer visible. Most  note worthy is the peak $\approx$ 1049s which is no longer visible. 

The approach that has been considered for this purpose is unlike the current BATHOTPIX\cite{r1} algorithm in NASA's Heasoft\cite{r4} Software package which considers the behavior of a particular pixel w.r.t other pixels in the detector plane and completely ignores the behavior of such a pixel over long intervals of time. For example, a pixel which does falls into the hottest pixel category at a particular instant of time might not fall into the same category at a different instant, similarly, a pixel falling in the centermost\cite{r1} portion of the histogram at a particular instant of time, might not belong to a similar category at another instant of time. Considering the temporal variation of counts in the pixels is of utmost necessity if data from noisy pixels are to be eliminated.

\section*{Acknowledgements}
This research was supported by Tata Institute of Fundamental Research (TIFR), Mumbai and Goddard Space Flight Center. I would like to thank Prof. A. R. Rao from Department of Astronomy and Astrophysics, TIFR who provided insight and expertise that greatly assisted the research.

\end{document}